\documentclass[%
 reprint,
 amsmath,amssymb,
 aps,
]{revtex4-2}

\usepackage{graphicx}% Include figure files
\usepackage{dcolumn}% Align table columns on decimal point
\usepackage{bm}% bold math
\usepackage{color,soul}
\usepackage{amsmath}
\usepackage[utf8]{inputenc}
\usepackage[english]{babel}
\usepackage{setspace}

\usepackage{chngcntr}
\counterwithout{figure}{section}
\counterwithout{equation}{section}
\begin{document}

%\preprint{APS/123-QED}

\title{Electrically driven programmable phase-change meta-switch reaching 80\% efficiency}
%\thanks{A footnote to the article title}%

\author{Sajjad Abdollahramezani$^{1}$}
\author{Omid Hemmatyar$^{1}$}
\author{Mohammad Taghinejad$^{1}$}
\author{Hossein Taghinejad$^{1}$}
\author{Alex Krasnok$^{2}$}
\author{Ali A. Eftekhar$^{1}$}
\author{Christian Teichrib$^{6}$}
\author{Sanchit Deshmukh$^{3}$}
\author{Mostafa El-Sayed$^{4,5}$}
\author{Eric Pop$^{3}$}
\author{Matthias Wuttig$^{6}$}
\author{Andrea Al{\`u}$^{2, 7}$}
\author{Wenshan Cai$^{1,8}$}
\author{Ali Adibi$^{1}$}

\email{ali.adibi@ece.gatech.edu}

\affiliation{$^{1}$School of Electrical and Computer Engineering, Georgia Institute of Technology, 778 Atlantic Drive NW, Atlanta, Georgia 30332-0250, United States}
\affiliation{$^{2}$Photonics Initiative, Advanced Science Research Center, City University of New York, New York, NY 10031, United States}
\affiliation{$^{3}$Department of Electrical Engineering, Department of Materials Science and Engineering, Precourt Institute for Energy, Stanford University, Stanford, California 94305, United States}
\affiliation{$^{4}$Laser Dynamics Laboratory, School of Chemistry and Biochemistry, Georgia Institute of Technology, Atlanta, Georgia, 30332-0400, United States}
\affiliation{$^{5}$George W. Woodruff School of Mechanical Engineering, Georgia Institute of Technology, Atlanta, Georgia, 30332-0405, United States}
\affiliation{$^{6}$Physikalisches Institut IA, RWTH Aachen, Sommerfeldstrasse 14, 52074 Aachen, Germany}
\affiliation{$^{7}$Physics Program, Graduate Center, City University of New York, New York, NY 10016, United States}
\affiliation{$^{8}$School of Materials Science and Engineering, Georgia Institute of Technology, 801 Ferst Drive NW, Atlanta, Georgia 30332-0295, United States}

\date{\today}

% \begin{spacing}{2.5}

% \begin{large}

\begin{abstract}
Despite recent advances in active metaoptics, efficient wide dynamic range combined with high-speed reconfigurable solutions is still elusive. Phase-change materials (PCMs) offer a compelling platform for metasurface optical elements, owing to the large index contrast and fast yet stable phase transition properties. Here, we experimentally demonstrate an in situ electrically driven reprogrammable metasurface by harnessing the unique properties of a phase-change chalcogenide alloy, Ge$_{2}$Sb$_{2}$Te$_{5}$ (GST), in order to realize non-volatile, reversible, multilevel, fast, and pronounced optical modulation in the near-infrared spectral range. Co-optimized through a multiphysics analysis, we integrate an efficient heterostructure resistive microheater that indirectly heats and transforms the embedded GST film without compromising the optical performance of the metasurface even after several reversible phase transitions. A hybrid plasmonic-PCM meta-switch with a record eleven-fold change in the reflectance (an absolute reflectance contrast reaching 80\%), unprecedented quasi-continuous spectral tuning over 250 nm, and operation speed that can potentially reach a few kHz is presented. Our work represents a significant step towards the development of fully integrable dynamic metasurfaces and their potential for beamforming applications.
\end{abstract}


\maketitle

%\tableofcontents

\noindent\textbf{Introduction}

\noindent Optical metasurfaces, planar devices comprising densely arranged (a-)periodic arrays of patterned structures, i.e., ``meta-atoms'', extend most functionalities realized by conventional bulky optical devices \cite{lin2014dielectric,arbabi2015subwavelength,rubin2019matrix,chen2018broadband,abdollahramezani2020meta,silva2014performing}. 
% owing to their constituent subwavelength inclusions that can spectrally, spatially, and/or temporally manipulate the optical properties (i.e., amplitude, phase, polarization, and dispersion) of an electromagnetic wave.
% To date, tremendous efforts have been devoted to realization of plasmonic and high-index dielectric metasurfaces to enable on-demand applications in lensing \cite{lin2014dielectric,arbabi2015subwavelength}, imaging \cite{rubin2019matrix,chen2018broadband}, and computing \cite{abdollahramezani2020meta,zangeneh2020analogue}. Due to the static nature of their constitutive inclusions, optical properties of such passive metasurfaces remain fixed upon fabrication, hindering the real-time modulation of light-matter interactions \cite{chen2016review,kuznetsov2016optically,dingrecent}.
Considering recent advancements in nanotechnology, developing high-performance reconfigurable metadevices, far surpassing the performance of their currently passive counterparts, would revolutionize modern optical technologies. 
Thus far, various tuning mechanisms benefiting from thermal, mechanical, electrical, chemical, and optical stimulation of dynamically tunable media, such as stretchable substrates, liquid crystals, graphene, doped semiconductors, and transparent conductive oxides, have been explored \cite{makarov2015tuning,huang2016gate, hail2019optical,zheludev2016reconfigurable,lewi2017ultrawide,shaltout2019spatiotemporal,kim2019phase,taghinejad2019all}. Despite impressive progress in the realization of tunable metasurfaces, existing demonstrations suffer from limitations including relatively weak optical modulation strength (due to low quality-factor (Q) nanoantennae), low optical performance (imposed by excessive material losses), low-speed modulation (limited by the intrinsic properties of tunable materials), and/or challenging manufacturing (on account of non-complementary metal oxide semiconductor (CMOS)-friendly fabrication processes). In this regard, new efficient material platforms improving the dynamic range of amplitude and phase modulations, facilitating pixel-level programming, and reducing static power consumption for the next-generation adaptive functional systems are in great demand. 

Phase-change materials (PCMs) with optical properties (e.g., refractive index) that are remarkably modified due to a pronounced change of properties upon crystallization attributed to the metavalent bonding in the crystalline phase \cite{kooi2020chalcogenides} can mitigate such a challenge \cite{feldmann2019all,zhang2019broadband,dong2019wide,delaney2020new,zheng2020nonvolatile,delaney2021non,shalaginov2021reconfigurable,wang2016optically}. Amongst existing PCMs, archetypal compound germanium antimony telluride (Ge$_2$Sb$_2$Te$_5$ or GST for short) has been vastly exploited in commercial rewritable optical disk storage technology and phase-change electronic memory applications exhibits attractive intrinsic features including non-volatility (long retention time of at least 10 years), ultrafast switching speed (10s-100s of ns), high switching robustness (potentially up to 10$^{12}$ cycles), considerable scalability (down to nanometer-scale lengths), low energy transition (down to a few aJ/nm$^3$), compatibility with CMOS processes, and good thermal stability, among others \cite{wuttig2007phase,dong2018tunable,wuttig2017phase,tian2019active,abdollahramezani2020tunable,michel2019advanced,gholipour2018phase}. Given the unique optical and electrical properties of GST, recently, significant attention has been paid toward the implementation of reconfigurable metadevices based on these properties \cite{tittl2015switchable,leitis2020all,tittl2015switchable,leitis2020all,abdollahramezani2021dynamic,de2018nonvolatile,yin2017beam,liu2020rewritable,qu2018thermal}.

To date, the dynamic tuning of phase-change metasurfaces has been entangled with active switching of PCMs between the amorphous and crystalline state using thermal-conduction annealing \cite{abdollahramezani2021dynamic} or laser pulses excitation \cite{li2016reversible}. While the former limits the device performance to one-way amorphous-to-crystalline conversion, both approaches necessitate the use of bulky external apparatus, i.e., a heating stage or an ultrafast laser. Inspired by the technological developments in mature phase-change random access memories, electrical threshold switching holds the promise to precisely control the states of the PCM in individual meta-atoms through a cross-bar array architecture \cite{kuzum2012nanoelectronic}. However, this scheme places stringent constraints on the targeted optical performance of the phase-change metasurface due to (i) interference of lossy metal wires with free-space light incident on the subwavelength meta-atoms, and (ii) formation of crystallization filamentation as a direct current path through PCM that prevents uniform phase transition of the whole PCM volume in meta-atoms. To meet these challenges, we leverage in situ electrical Joule heating using an optimized microheater design that indirectly actuates the PCM element of our metasurface. Our demonstration largely surpasses the state-of-the-art alternatives for reconfigurable metasurfaces in four important dimensions. First, the heterostructure metadevice platform formed by the integration of a robust microheater underneath the metasurface enables uniform electrothermal phase conversion without adding excessive dissipative loss to the optical device. Reaching 80\% optical efficiency, our platform outperforms the recently developed reflector-absorber PCM-switches \cite{zhang2021electrically,wang2021electrical}. In addition, our architecture significantly reduces the deformation of the meta-atoms caused by inevitable heating of the alternative resistive microheaters that use plasmonic materials like silver \cite{wang2021electrical} with low melting temperature. Second, our electrically driven platform enables achieving multiple non-volatile intermediate PCM states (between amorphous and crystalline) in a repeatable fashion to realize multi-state reconfigurable metasurfaces necessary for adaptive flat optics. Third, we use technologically mature GST that provides highest index contrast among all PCMs at the near-infrared (near-IR) spectral range. This allows us to fairly decrease the thickness of the PCM layer without compromising the optical performance. The application of a thin PCM layer is necessary for the realization of a repeatable and reliable melt-quenching process while avoiding elemental segregation as a typical failure mechanism of PCMs \cite{wuttig2017phase,abdollahramezani2020tunable}. Furthermore, orders-of-magnitude faster registration of intermediate states with lower dynamic power can be performed in GST in comparison to the recently emerged phase-change alloys such as Ge$_{2}$Sb$_{2}$Se$_{4}$Te$_{1}$ (GSST) \cite{zhang2021electrically}. Finally, the fundamental modes of the metasurface on account of the near-field interaction of incident light with plasmonic elements exhibit good modal overlap with the GST film, which facilitates the manipulation of the optical scattering with a wide-dynamic range. The successful operation of our electrically driven programmable metasurface owes to a judicious co-optimization of a multiphysics model taking into account the extreme electrical, thermal, and optical properties of the contributing materials. Being the first demonstration of a fully reversible reconfigurable GST-based metasurface with multiple intermediate states and a large tuning range, our platform has the potential for major applications in several fields including imaging, sensing, neural computing, and quantum information processing.

% \section{Results and Discussions}

\begin{figure*}
\centering
\includegraphics[width=.8\linewidth, trim={0cm 0cm 0cm 0cm},clip]{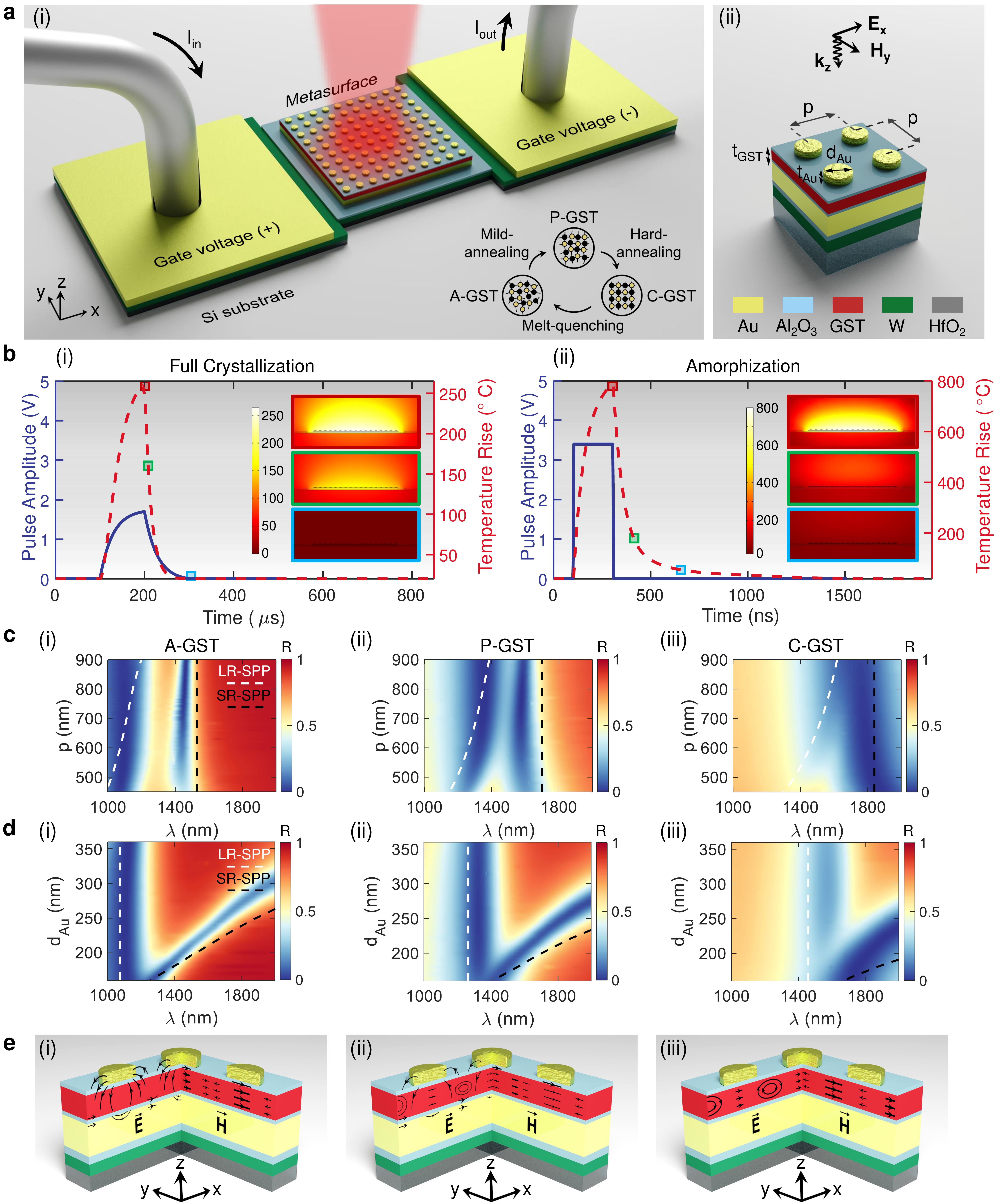}
\caption{ \textbf{Operation principle of an electrically reconfigurable heterostructure metadevice.} (\textbf{a}) (i) A perspective view of the heterostructure consisting a resistive microheater integrated with a phase-change metasurface. The microheater is constructed by deposition of a 12 $\mu$m$\times$12 $\mu$m resistive patch of W connecting two 100 $\mu$m$\times$100 $\mu$m Au probing pads on top of a $\textrm{HfO}_{2}$-coated Si substrate. The metasurface is formed by multilayer deposition of an optically-thick Au backreflector, a GST film encapsulated between two equally-thick protective layers of Al$_2$O$_3$, and a 2D array of Au nanodisks. Inset: a generic scheme of atomic distribution of partial crystalline GST (P-GST), full crystalline GST (C-GST), and amorphous GST (A-GST) after mild-annealing by applying a long low-intensity pulse, hard-annealing by applying a long medium-intensity pulse, and melt-quenching by applying a short high-intensity pulse. (ii) Cross section view of the heterostructure metadevice. The metasurface has the same period $p$ in both $x$ and $y$ directions. The diameter of the Au nanodisk is denoted by $d_{\textrm{Au}}$ while $t_{\textrm{m}}$ represents the thickness of material ``m''. (\textbf{b}) Real-time voltage of the applied ``set'' and ``reset'' electrical pulses (solid blue lines) and the corresponding temperature responses (dashed red lines) in the center of the GST film for (i) full crystallization and (ii) amorphization processes. Inset: simulated temperature distributions at the cross section of the phase-change metasurface at the time marked by the color-coded markers. (\textbf{c,d}) Simulated 2D reflectance maps for the metasurface as functions of (\textbf{c}) $p$ and (\textbf{d}) $d_{\textrm{Au}}$ with (i) A-GST, (ii) P-GST, and (iii) C-GST. The coupling of the incident light to the LR-SPP mode is displayed by the white dashed line while the black dashed line represents the evolution of the SR-SPP mode. $t_{\textrm{GST}} = 40$ nm, $t_{\textrm{Al$_2$O$_3$}} = 10$ nm, and $t_{\textrm{Au}} = 35$ nm are fixed while $d_{\textrm{Au}} = 200$ nm and $p = 600$ nm are chosen in (\textbf{c}) and (\textbf{d}), respectively. (\textbf{e}) The electric and magnetic field vector distributions for (i) SR-SPP, (ii) the hybrid mode, and (iii) LR-SPP within the $x-z$ and $y-z$ planes of a meta-atom, respectively.} 
\label{fig:Fig_1}
\end{figure*}

\noindent\textbf{Electrothermal analysis of the integrated heterostructure metadevice}

\noindent A schematic view of the reconfigurable heterostructure metadevice driven by in situ electrical pulses is represented in Fig.~\ref{fig:Fig_1}a(i). The configuration of the microheater is carefully chosen to meet the design specifications within the limitation of testing equipment. It consists of a 12$\times$12 $\mu$m$^2$ square of 50-nm-thick tungsten (W) layer connected to the top metasurface with a 20-nm-thick layer of alumina (Al$_2$O$_3$) and isolated from the silicon (Si) substrate, as a good heat sink, with a 100-nm-thick hafnia ($\textrm{HfO}_{2}$) film. The microheater is in contact with two 100-$\mu$m-wide gold (Au) pads to facilitate the engagement with high-frequency probes or wire bonding to an external board. We chose W for the microheater material due to its highest melting point, good thermal conductivity, moderate resistivity, and low thermally activated diffusion. While high thermal conductivity of Al$_2$O$_3$, in comparison to all existing oxides, significantly facilitates heat exchange between the microheater and the metasurface, a thick-enough $\textrm{HfO}_{2}$ layer helps preservation of the generated heat for GST phase change to keep the electrical power consumption low. The fabrication processes are detailed in Methods and Supplementary Fig.~S1.

In order to study the performance of the miniaturized heater in terms of switching speed, heating/cooling rate, temperature uniformity, and energy efficiency, the real-time temperature distribution of the heterostructure metadevice in response to two kinds of electrical pulses with different temporal profiles (i.e., ``set'' and ``reset'') were calculated. As shown in {Fig.~\ref{fig:Fig_1}}b(i), the low-voltage set pulse (with 200-$\mu$s-long double exponential waveform and a peak voltage of 1.7 V) heats amorphous GST above the crystallization temperature ($\sim$160 $^{\circ}$C \cite{yamada1991rapid}) for a sufficiently long time to ensure adequate nucleation and formation of crystalline islands. On the other hand, a high-voltage pulse (with 200-ns-long rectangular waveform and a peak voltage of 3.4 V) featuring very short leading/trailing edge ($\sim$5 ns) rapidly increases the temperature of GST above the melting temperature ($\sim$630 $^{\circ}$C \cite{yamada1991rapid}) followed by quenching such that GST solidifies in the amorphous state (see {Fig.~\ref{fig:Fig_1}}b(ii)). Considering the enlarged size of the microheater (one order of magnitude larger than those previously studied \cite{taghinejad2020ito,zhang2019miniature,zheng2020nonvolatile,rios2021multi}), uniform heat generation over the whole volume of the GST film should be carefully addressed to ensure reliable and repeatable optical performance. The simulated two-dimensional (2D) temperature map in Fig.~\ref{fig:Fig_1}b(i) indicates that at the end of the set pulse, the temperature difference between the center and the two ends of the metasurface is $<$ 12\%. This is uniform enough to guarantee repeatable conversion of the GST film though it can be improved by increasing the clearance between the metasurface and the end of large probing pads acting as a heat sink (see Supplementary Note 1 and Fig.~S2). The data cut along the $x$ axis reveals that, thanks to the thinness of the GST film, the temperature gradient along the out-of-plane direction is negligible ($<$ 0.2\%). These features are essential to precisely register multiple reversible intermediate phases to the GST film and enabling reprogrammable multifunctional metasurfaces, a key attribute of our work.

The realization of sufficiently fast cooling rate is a grand challenge towards the amorphization process of PCMs. If the thermal characteristic of the microheater is modeled as a first-order system comprising a parallel thermal resistance (R$_{t}$) and thermal capacitance (C$_{t}$), the cooling rate follows an exponential relation with a time constant of $\tau$ = R$_{t}$C$_{t}$. R$_{t}$ and C$_{t}$ depend on the length and the width of the microheater as well as the thermal properties of the ambient. Therefore, to speed up the metadevice transient response, an optimized selection of material and geometry for both the microheater and the surrounding medium is indispensable (see Methods and Supplementary Note 1). As depicted by electrothermal simulations in {Fig.~\ref{fig:Fig_1}}b(ii), the heterostructure with an elevated temperature of 790 $^{\circ}$C at the end of the reset pulse is cooled down with the rate of $\sim$10 $^{\circ}$C/ns to $<$ 480 $^{\circ}$C and $\sim$6 $^{\circ}$C/ns to $<$ 160 $^{\circ}$C, which is higher than typical $1$ $^{\circ}$C/ns melt-quenching criterion \cite{wuttig2017phase,abdollahramezani2020tunable}.

\noindent\textbf{Design of the phase-change metasurface}

\noindent We reveal the high potential of phase-change metasurfaces in engineering the optical scattering by exploring the near-field light-matter interaction mechanisms within a hybrid plasmonic-PCM meta-atom. Through a set of theoretical calculations and full-wave electromagnetic simulations (see Methods and Supplementary Note 2 for details), we investigate in particular the evolution of governing surface plasmon polariton (SPP) modes and the intricate coupling processes between them upon the phase transition of GST from amorphous (A-GST for short) to partial crystalline (P-GST for short) and ultimately to full crystalline (C-GST for short).

The 10.5$\times$10.5 $\mu$m$^2$ phase-change metasurface comprises 35-nm-thick Au nanodisks separated from a 80-nm-thick Au backreflector by a 40-nm-thick blanket film of GST sandwitched between two 10-nm-thick layers of Al$_2$O$_3$ (see {Fig.~\ref{fig:Fig_1}}a(ii); also, see Methods for fabrications details). Two Al$_2$O$_3$ layers prevent heating-induced oxidation of the GST film and diffusion of the noble metal into GST during the heating process. In the case of A-GST with low intrinsic loss (see ellipsometric measurements in Supplementary Fig.~S3), the hybrid plasmonic-PCM meta-atom supports two spectrally distant SPP modes, namely long-range SPP (LR-SPP) and short-range SPP (SR-SPP) (see Figs.~\ref{fig:Fig_1}c(i) and \ref{fig:Fig_1}d(i)) \cite{barnard2008spectral}. The theoretically calculated spectral locations of the former and the latter are displayed by white and black dashed lines, respectively. Figures~\ref{fig:Fig_1}c(ii) and \ref{fig:Fig_1}d(ii) depict that partial crystallization of GST draws the slightly broadened modes to the center of the telecom spectral window, i.e., 1260-1650 nm, where they fairly overlap. By fully converting the state of GST using electrical Joule heating, a significant index contrast can be observed (see Supplementary Fig.~S3), which further broadens and dampens the existing resonance modes, as shown in Figs.~\ref{fig:Fig_1}c(iii) and \ref{fig:Fig_1}d(iii). The spatial characteristics of the LR-SPP and SR-SPP modes due to the excitation of SPPs at the infinite interface of the Au backreflector and the bottom Al$_2$O$_3$ layer and that at the individual nanodisk and the top Al$_2$O$_3$ layer are schematically illustrated in Figs.~\ref{fig:Fig_1}e(i) and \ref{fig:Fig_1}e(iii). The electric and magnetic flowlines in Fig.~\ref{fig:Fig_1}e(ii) reveal the existence of a hybrid mode due to the overlap between the localized SR-SPP and distributed LR-SPP modes.
The rich physical properties and distinct characteristics of governing modes coupled to the available state of GST offer a good degree of freedom for realization of multifunctional metasurfaces. Particularly, the evolution of the governing mode from SR-SPP in A-GST (with higher plasmonic and lower photonic loss) to overlapped SR-SPP/LR-SPP in P-GST (with balanced plasmonic and photonic losses) and finally to LR-SPP in C-GST (with lower plasmonic and higher photonic loss) facilitates manipulation of both amplitude and phase properties of light (see Supplementary Note 3, Figs.~S4 and S5) in the telecom wavelengths.

\begin{figure*}
\centering
\includegraphics[page=1,width=.8\linewidth, trim={0cm 0cm 0cm 0cm},clip]{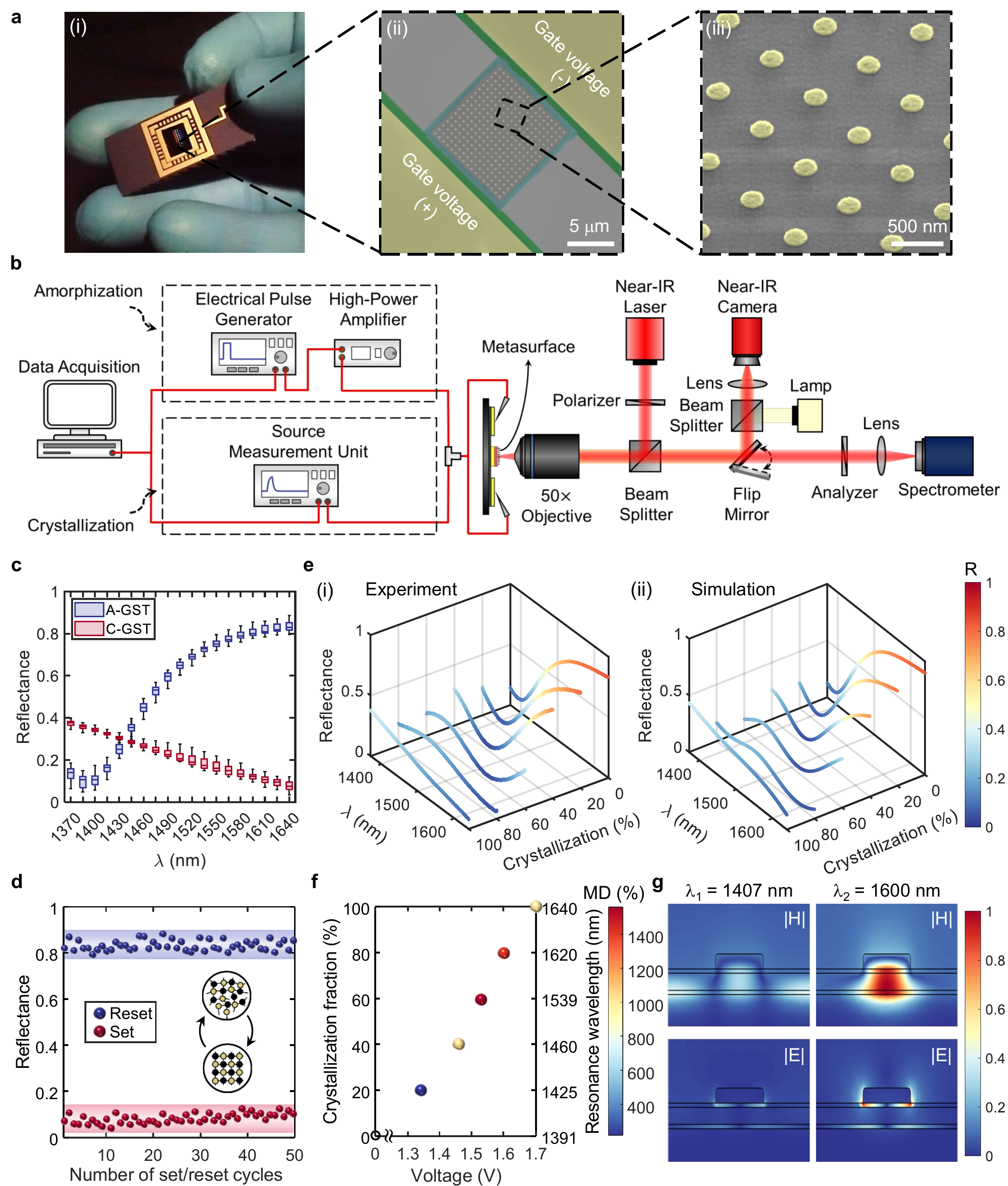}
\caption{ \textbf{Experimental characterization of the electrically programmable meta-switch.} (\textbf{a}) (i) Image of the fabricated sample mounted on a ceramic chip carrier, (ii) tilted false-colored SEM image of the meta-switch comprising the microheater and the phase-change metasurface, and (iii) the magnified bird's eye view of the meta-atom array. (\textbf{b}) Schematic of the experimental near-IR reflectometry setup coupled to signal generators for co-located electrical programming and optical characterization of fabricated metadevices. (\textbf{c}) Demonstration of the binary operation of the meta-switch: statistical distribution of change in measured reflectance over 50 consecutive cycles of crystallization (red boxes) and amorphization (blue boxes) for 15 equal-distant wavelengths. The lower/upper quartile and the median of the collected data are represented by color-coded lines while the minimum/maximum extent are displayed with black line. (\textbf{d}) Cyclability plot of the optical reflectance of the meta-switch during multiple electrical set (red dots leading to C-GST) and reset (blue dots leading to A-GST) pulses. The 95\% confidence intervals (shaded areas) of $\pm$1\% and $\pm$7.5\% are measured for the reflective and absorptive states, respectively. (\textbf{e}) Demonstration of the multi-state operation of the meta-switch: (i) measured and (ii) simulated color-coded reflectance spectra of the programmed meta-switch with A-GST (i.e., 0\% crystallization fraction), C-GST (i.e., 100\% crystallization fraction), and 4 accessed intermediate states (with 20\% crystallization steps). (\textbf{f}) Correlation of the modulation depth (MD), undergoing approximate crystallization fraction, and fundamental resonance wavelength shift with the applied pulse voltage. (\textbf{g}) Inspection of the normalized magnetic field magnitude and electric field magnitude at the resonance wavelengths of the meta-switch with 80\%-crystalline GST (in panel (\textbf{e})(ii)) in the $x-z$ plane. At $\lambda_1$, the extended interface-bounded LR-SPP mode is excited due to the grating effect of the nanodisk array whereas at $\lambda_2$, the SR-SPP mode features a strong localized magnetic filed underneath the individual nanodisk due to mirrored dipoles. The structural parameters of the studied metasurface are $p = 600$ nm, $d_{\textrm{Au}} = 190$ nm, $t_{\textrm{Au}} = 35$ nm, $t_{\textrm{GST}} = 40$ nm, and $t_{\textrm{Al$_2$O$_3$}} = 10$ nm.}
\label{fig:Fig_2}
\end{figure*}

\noindent\textbf{Dynamic multistate meta-switch characterization}

\noindent We leverage the rich nature of the SPP modes and large index contrast of GST for the demonstration of an electrically driven multispectral meta-switch. Figure~\ref{fig:Fig_2}a(i) shows the fabricated sample (see Methods for details) mounted on a ceramic carrier chip to facilitate packaging and external computerized bias control (not shown here). A scanning electron microscope (SEM) image of the heterostructure metadevice and a bird's-eye view of the metasurface consisting of an array of 17$\times$17 identical meta-atoms are depicted in Figs.~\ref{fig:Fig_2}a(ii) and \ref{fig:Fig_2}a(iii), respectively. We use a home-built linear reflectometery setup coupled to two signal generators for co-located electrical excitation and optical measurement of fabricated samples (see Fig.~\ref{fig:Fig_2}b and Methods for details). 

The structural design parameters of the metasurface are judiciously chosen to keep fundamental resonances of the meta-switch at the two extreme cases of GST far apart, which guarantees high modulation depth over a wide spectral bandwidth. To corroborate the design strategy, the statistical distribution of experimentally measured reflectance spectra over 50 cycles of crystallization-amorphization is displayed in Fig.~\ref{fig:Fig_2}c (see Supplementary Fig.~S6 for the detailed reflectance spectra of all cycles). The narrow boxes reveal the slight deviation of the first and the third quartiles from the median of sampled data for 15 discrete wavelengths in the telecom range. It is evident that the resonance wavelength of the meta-switch red shifts from  1391  nm  with  A-GST  to  1640  nm  with  C-GST. Upon this transition, an average absolute (${\Delta}R = |R_\textrm{A-GST}-R_\textrm{C-GST}|$) and relative modulation depth (${\Delta}R/R_\textrm{C-GST}$) over 75\% and $1000\%$ are achieved at 1640 nm, respectively. More importantly, with average 82\% reflectance in the reflective state, our platform surpasses the state-of-the-art electrically tunable PCM meta-switches \cite{zhang2021electrically,wang2021electrical}. For more clarity, time-dependent traces of the change in the reflectance at 1640 nm during consecutive cycles of switching are depicted in Fig.~\ref{fig:Fig_2}d. The measured 95\% confidence intervals (shaded areas) of $\pm$1\% and $\pm$7.5\% for the reflective and absorptive state, respectively, verify the highly reproducible switching process. Such consistent characteristics is also verified through confocal Raman microscopy by studying the micro-Raman scattering from the meta-switch under test (see Supplementary Fig.~S7). The non-deterministic behavior of the absorptive and reflective state stems from formation of non-homogeneous crystalline regions during heating and stochastic recrystallization of small islands in GST during the quenching process, respectively \cite{yamada1991rapid,wuttig2007phase,kuzum2012nanoelectronic}. Considering the fast relaxation time of GST and small thermal time constant and temperature nonuniformity of the heterostrucutre metadevice (as verified by electro-thermal analysis in Fig.~\ref{fig:Fig_1}b(i)), we expect an operation speed of a few kHz with order-of-magnitude lower operational voltages compared to state-of-the-art electro-mechanical optical metasurfaces \cite{arbabi2018mems}.

Besides the binary-level switching, distinctive and stable intermediate phases of GST, in virtue of its giant index contrast, non-volatile, and nucleation-dominant characteristics, hold the promise for multi-state switching operation. Considering the good thermal uniformity across the microheater, precise control of the crystalline fraction of GST, through formation of critical nuclei and their subsequent growth, can be realized by applying a customized electrical pulse. We explore this capability by programming the meta-switch with consecutive fixed length pulses with different voltages. The measured reflectance of the meta-switch for A-GST, C-GST, and 4 accessed intermediate phases of GST (see Fig.~\ref{fig:Fig_2}e(i)) show quasi-continuous tunning of the fundamental resonances (i.e., LR-SPP and SR-SPP) from $\sim$1391 nm to $\sim$1640 nm. With such record optical contrast, unprecedented ultrawide spectral tuning range, and potential fast switching operation, our platform outperforms many existing works relying on electro-optical, electro-mechanical, and thermo-optic effects \cite{shaltout2019spatiotemporal}.

To quantitatively analyze the crystallization kinetics upon electrical pulse excitation, we compare the measured reflectance spectra with simulated ones for different crystallization fractions of GST, whose optical properties are approximated using an effective medium theory (see Supplementary Note 4). As shown in Fig.~\ref{fig:Fig_2}e, a good agreement is observed between the color-coded experimental measurements and simulated results from intermediate states with $\sim$20\% crystallization steps. Figure~\ref{fig:Fig_2}f depicts the correlation between the measured applied pulse voltage, resonance wavelength of the meta-switch, and approximated crystallization fraction. Evidently, a wide spectral tuning range is achieved upon multi-state conversion of GST using electrical pulses with small voltages. We further quantitatively investigate the crystallization fractions of GST in different intermediate states as a function of the induced temperature (see Supplementary Note 1 and Fig.~S2).

To study the physical mechanism behind the operation of the meta-switch, we calculate the electromagnetic field distribution at the two dips of the reflectance spectrum for the intermediate case with 80\% crystallization fraction (see Fig.~\ref{fig:Fig_2}e(ii)). The field profiles in the $x-z$ cross section of a meta-atom in Fig.~\ref{fig:Fig_2}g show excitation of the SR-SPP mode and LR-SPP mode for $\lambda_{1} = 1407$ nm and $\lambda_{2} = 1600$ nm, respectively. The two SPP modes do not exhibit the same degree of localization and enhancement. For the former case, the magnetic field is distributed along the interface of the Au backreflector and the bottom Al$_2$O$_3$. For the latter, the magnetic field is strongly enhanced underneath the nanodisk due to the anti-symmetric current distribution in the two Au parts. The electric field magnitude profile in Fig.~\ref{fig:Fig_2}g (and flowlines of the Poynting vector in Supplementary Fig. S8a)  implies that a good portion of the incident energy is dissipated after funneling of the incident wave into the lossy GST film at $\lambda_{1}$. In contrast, the coupling of accumulated charges at both lateral end-faces of the nanodisk can form a pronounced electric dipole resonance at $\lambda_{2}$. Such a strong resonance fairly traps the major energy of incident light near the nanodisk that is dissipated due to the lossy nature of Au (also see flowlines of the Poynting vector in Supplementary Fig.~S8b).

\begin{figure*}
\includegraphics[page=1,width=1\linewidth, trim={0cm 0cm 0cm 0cm},clip]{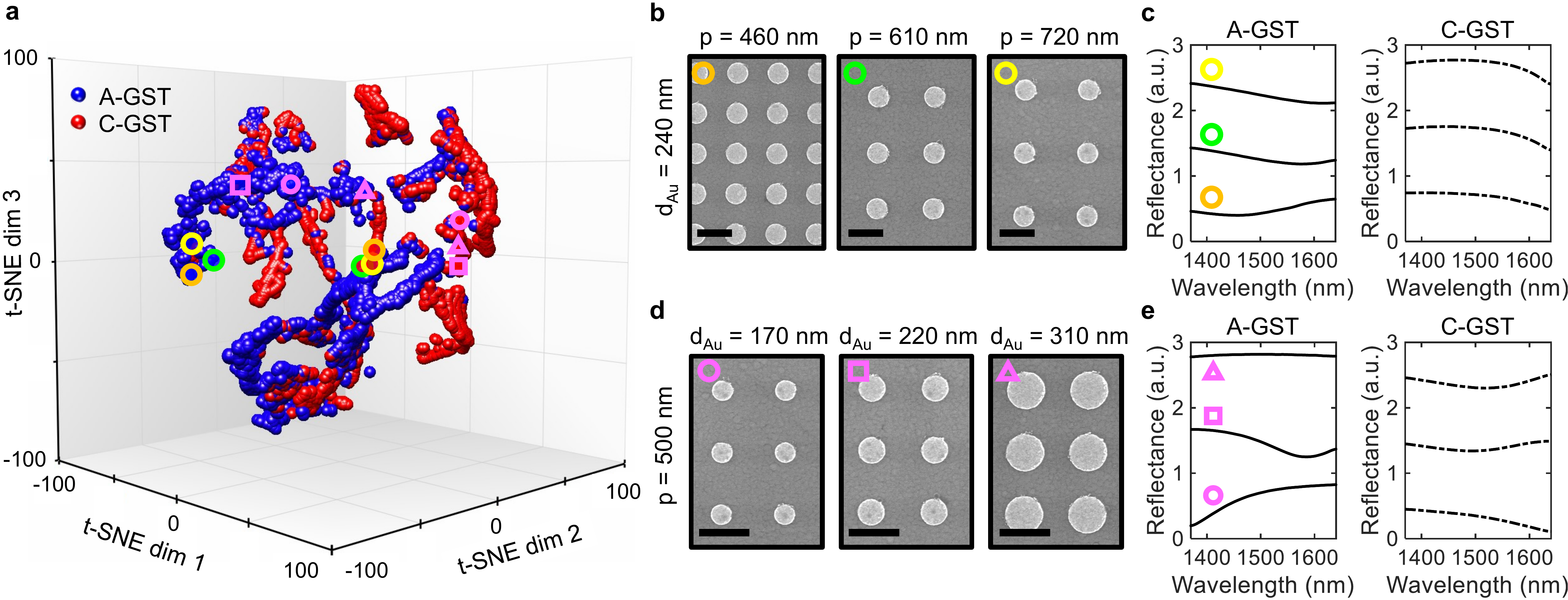}
\caption{\textbf{Performance analysis of the reconfigurable meta-switch using machine learning.} (\textbf{a}) Unsupervised 3D nonlinear embedding of the simulated reflectance spectra of the phase-change metasurface in Fig.~\ref{fig:Fig_2}a with different structural parameters (i.e., $p$, $d_\textrm{Au}$, $t_\textrm{GST}$, and $t_\textrm{Al$_2$O$_3$}$). The embeddings corresponding to A-GST and C-GST are depicted as blue and red points, respectively. The marked data are the low-dimensional representation of experimental results from the fabricated metasurfaces with the following parameters: orange circle ($d_\textrm{Au}=240$ nm, $p=460$ nm), green circle ($d_\textrm{Au}=240$ nm, $p=610$ nm), yellow circle ($d_\textrm{Au}=240$ nm, $p=720$ nm), magenta circle ($d_\textrm{Au}=170$ nm, $p=500$ nm), magenta square ($d_\textrm{Au}=220$ nm, $p=500$ nm), and magenta triangle ($d_\textrm{Au}=310$ nm, $p=500$ nm). Other structural parameters are $t_{\textrm{Au}} = 35$ nm, $t_{\textrm{GST}} = 40$ nm, and $t_{\textrm{Al$_2$O$_3$}} = 10$ nm.
(\textbf{b,d}) SEM images of fabricated metasurfaces with the experimental data marked in (\textbf{a}) and their corresponding reflectance responses in (\textbf{c,e}), respectively. The scale bars in all SEM images indicate 400 nm. The reflectance curves in (\textbf{c,e}) are shifted along the vertical axis for the sake of clarity. The embedding of data corresponding to metasurfaces with the same $d_\textrm{Au}$ but different $p$ (\textbf{b,c}) are close to each other, while those with the same $p$ but different $d_\textrm{Au}$ (\textbf{d,e}) are apart from each other as shown in (\textbf{a}).}
\label{fig:Fig_3}
\end{figure*}

\noindent\textbf{Performance analysis using machine learning}

\noindent Beyond formal modeling of any physical phenomenon, exploratory analysis with diagrammatic representations is a powerful tool that helps inferring by visualization of the data. The key concept is to form an easy-to-interpret low-dimensional representation of the structured data with the end goal of unveiling data points with unusual attributes, demystifying the underlying connections, and revealing the governing patterns. In this regard, to gain an intuitive understanding of the overall performance of the meta-switch, without relying on the apriori knowledge, we leverage data visualization using an unsupervised machine learning approach. This method transforms a set of high-dimensional data sets (e.g., high-resolution reflectance spectra in both A-GST and C-GST states of the meta-switch) into low-dimensional maps while preserving necessary information (e.g., nature of the governing mode) \cite{bengio2013representation}. Among several options \cite{gisbrecht2015data, ma2020deep}, we leverage a nonlinear dimensionality reduction technique called t-distributed stochastic neighbor embedding (t-SNE) extensively used for data exploration and visualization of high-dimensional data in image processing \cite{maaten2008visualizing}. The t-SNE algorithm aims to match neighbors in a higher-dimensional space to a lower-dimensional one by measuring the similarity between pairs of variables. It then optimizes these two similarity measures based on a predefined cost function. Upon applying to a high-dimensional but well-clustered data set, t-SNE tends to generate a visual embedding with distinctly isolated clusters.

{Figure~\ref{fig:Fig_3}}a represents three-dimensional (3D) embeddings of reflectance responses of the metasurface in {Fig.~\ref{fig:Fig_1}}a with different structural parameters for both A-GST and C-GST cases. Numerical simulations are carried out in the operational spectral range, i.e., 1370-1640 nm, for 3600 different metasurfaces with a wide range of randomly selected values for the structural parameters (i.e., $p$, $d_\textrm{Au}$, $t_\textrm{GST}$, and $t_{\textrm{Al}_2\textrm{O}_3}$) in both cases of A-GST and C-GST. Evident from {Fig.~\ref{fig:Fig_3}}a, two widely unfolded clusters corresponding to A-GST (blue points) and C-GST (red points) are extended over the 3D latent space. This implies that incorporation of GST in the meta-atom considerably spans the attainable responses not accessible through just variation of structural parameters with one GST state. Accordingly, the GST crystallization state can be considered as an ideal tuning knob to substantially modify the metadevice performance. Furthermore, the minimum overlap between these clusters suggests that metasurfaces with A-GST and C-GST are governed by modes with distinct natures. 
To elucidate the results, the reflectance spectra for metasurfaces with 2 specific sets of structural designs, i) $d_\textrm{Au} = 240$ nm with varying $p$, and ii) $p = 500$ nm with varying $d_\textrm{Au}$ (see corresponding SEM images in Figs.~\ref{fig:Fig_3}b and \ref{fig:Fig_3}c) are shown in Figs.~\ref{fig:Fig_3}d and \ref{fig:Fig_3}e, respectively. The responses are also included in the 3D dimensionality-reduced space in Fig.~\ref{fig:Fig_3}a using color-coded shapes. It is seen that for case (i), the corresponding color-coded circles are fairly centered in the 3D embeddings while for case (ii), the magenta shapes are distinguishably expanded over the 3D latent space. Looking into Figs.~\ref{fig:Fig_3}c and \ref{fig:Fig_3}e, the reflectance spectra for the former case, specifically for A-GST, evolve more rapidly than that for the latter case.

While the t-SNE algorithm provides a helpful visualization of the range of responses upon changing each design parameter, it is not straightforward to use it to compare the importance of the different design features in varying the optical response. Such information is crucial in various aspects: i) it provides valuable insight about the  robustness of switching operation against variation of each design parameter, ii) it can be used to devise non-uniform sampling of the overall range of different design parameters to form the (random) training dataset for considerably reducing the computation requirements, and iii) it identifies the most vulnerable parameters to the fabrication errors to help in customizing the optimal fabrication process. The shortcoming of t-SNE as well as other nonlinear dimensionality-reduction (sometimes known as feature transformation) algorithms is that the dimensionality-reduced variables (or bases) in the latent space are complex nonlinear combinations of all input (or design) parameters. To address this issue and enable the benefits of ranking the importance of design parameters, feature-selection algorithms can be used. We employ a supervised approach, namely the wrapped method \cite{guyon2003introduction}, utilizing a machine learning algorithm with the cost function properly defined to rank the most effective design parameters in achieving the maximum relative modulation depth in the spectral range of 1370-1640 nm. Starting from an empty feature subset, the algorithm sequentially adds each of the structural parameters as a candidate to the subset and performs cross-validation by repeatedly calculating the evaluation criterion until the stopping condition is reached.

By implementing the wrapping algorithm on the training dataset of our metasurface structures, we find that the most influential design parameters (in the high-to-low order) are $t_\textrm{GST}$, $d_\textrm{Au}$, $p$, and $t_{\textrm{Al}_2\textrm{O}_3}$.  Here, we assume the crystalline state of GST is fixed, and we find the same ranking for A-GST, P-GST, and C-GST. Note that this conclusion is made through the learning process without adding any physical apriori knowledge. The conclusions are supported by the properties of the electromagnetic modes of the structure. As unveiled in Fig.~S9a, variation of $t_\textrm{GST}$ significantly affects the spectral evolution of both SR-SPP and LR-SPP modes, where the anti-crossing behavior can occur due to pronounced coupling of these two modes. That is to say the enhancement of the induced magnetic dipole underneath the nanodisk and the coupling strength of the incident light to the in-plane wave are closely linked to this design parameter. In addition, comparison of Figs.~\ref{fig:Fig_1}c and \ref{fig:Fig_1}d infers the more sensitivity of the optical performance of the metasurface to the characteristics of SR-SPP, which is mainly influenced by changes in $d_\textrm{Au}$, than those of the LR-SPP, whose evolution is basically governed by changes in $p$. Figure~S9b reveals that variation of $t_\textrm{Al$_2$O$_3$}$ negligibly affects the overall optical performance of the metasurface validating the results from our feature-selection approach.

\noindent \textbf{Conclusion}

\noindent In summary, we demonstrated a chip-scale electrically driven phase-change metadevice through incorporation of a heterostructure microheater integrated with a hybrid plasmonic-PCM metasurface. A systematic design approach by leveraging a multiphysics electro-thermal study and an electromagnetic analysis of the metadevice was presented. Without comprising the optical performance, our introduced platform is capable of quasi-continuous, reversible, and non-volatile tuning of the fundamental modes of the meta-atom, i.e., LR-SPP and SR-SPP. An active meta-switch with an unprecedented optical efficiency reaching 80\%, unparalleled electrical modulation of the reflectance by more than eleven-fold, unique spectral tuning over 250 nm, and potential operation speed of a few kHz was experimentally demonstrated. We leveraged data-driven machine learning approaches to reveal the undeniable effect of GST on expanding the attainable response space of a reconfigurable metasurface and rank strategical structural parameters influencing the overall optical response of the meta-switch. Our findings can open new directions for reconfigurable flat optical devices manipulating the dominant properties of light for various applications including imaging, computing, and ranging.

% Experimental section

\noindent \textbf{Methods}

\noindent \textbf{Electrothermal simulations.}
The COMSOL Multiphysics software package is used for Joule heating simulations and consequently transient thermal behavior study of the electrically driven metasurface. A coupled multiphysics model including the Electric Currents module, for simulating the electrical current profile, and Heat Transfer in the Solid module, for calculating the heating exchange and temperature distribution, is employed. In the Heat Transfer module, infinite element domains are considered for the side boundaries of the constructed model. In addition, the top and bottom surfaces are given a convective heat flux boundary condition with ambient temperature of $T=20^{\circ}$ C. The parameters of contributed materials are taken from four-point probe measurements and existing experimental data in the literature (see Supplementary Note 1 for details).
% and standard COMSOL libraries, except for the electrical resistivity of W that obtained from four-point probe measurements of the fabricated sheet film.

\textbf{Numerical simulations.}
Full-wave simulations are carried out using the commercial finite element method (COMSOL Multiphysics) and verified by the finite integral technique software CST Microwave Studio. We model the meta-atom of the metasurface by considering periodic boundary conditions on the vertical sidewalls along the $x$ and $y$ directions. We apply perfectly matched layers at the truncated air boundary along the $z$ direction to avoid spurious back reflections. In our simulations, a broadband propagating plane wave perpendicularly (unless otherwise stated) is launched toward the metasurface from free space. A 2D monitor is used in the free space above the metasurface to record the reflected light amplitude. The refractive index of Al$_2$O$_3$ and GST used in the simulations are obtained through spectroscopic ellipsometry measurements (see Supplementary Note 4 and Fig.~S3). The optical constants of Au is obtained from experimental values reported in Ref. \cite{palik1998handbook}.

\noindent \textbf{Sample preparation.}
The electrically driven reprogrammable metasurface is implemented through a series of standard and customized fabrication processes (see the flow diagram in Supplementary Fig.~S1). We start with the atomic layer deposition (ALD) of a 100-nm-thick HfO$_{2}$ layer on a 500-$\mu$m-thick Si substrate to prevent direct contact between the microheater and probing pads and the base substrate. Next separate steps involve electron beam (e-beam) lithography to define the patterns of the microheater/probing pads followed by e-beam evaporation of a 50-nm-thick W layer/a 100-nm-thick Au layer and ultimately a lift-off process. Then, e-beam lithography is performed to define the aperture on the microheater where the metasurface is finally located. Sequential depositions of a 20-nm-thick Al$_2$O$_3$ layer by ALD, an 80-nm-thick Au layer by e-beam evaporation, and a 10-nm-thick Al$_2$O$_3$ layer by ALD are performed to fill the opening. It follows by the deposition of a 40 nm-thick GST layer from a stochiometric target in an RF-sputtering system and subsequent deposition of a 10-nm-thick Al$_2$O$_3$ as a capping layer in the ALD chamber. After the lift-off process, spin coating of a thin layer of polymethyl methacrylate (PMMA) is performed and Au nanodisk arrays are lithographically defined and formed by developing in a room-temperature methyl isobutyl ketone/isopropyl alcohol (MIBK/IPA) mixture. As the last step, e-beam evaporation of a 35-nm-thick Au layer is carried out followed by an overnight lift-off process. An ultrathin layer of Titanium (Ti) is used as the adhesion for Au.

\noindent\textbf{Material characterization.}
The complex refractive indices of A-GST and C-GST with different thicknesses are calculated using spectroscopic ellipsometery measurements with three different angles of incidence (50$^{\circ}$, 60$^{\circ}$, 70$^{\circ}$) over 350-2000 nm spectral range (see Supplementary Fig.~S3). Tauc-Lorentz and Cody-Lorentz \cite{shportko2008resonant} are chosen as fitting models with optical bandgap, oscillator width, Lorentz oscillator amplitude, resonance energy, and Urbach energy as fitting parameters. We observe that the thickness of GST uniformly shrinks by $\sim$5\% upon full crystallization (see Supplementary Fig.~S10), which is in agreement with the experimental results in the literature. The surface roughness measured by the atomic force microscopy (AFM) measurement (see Supplementary Fig.~S11) is used in our calculations. We benefit from confocal Raman microscopy to study the Raman scattering of the A-GST/C-GST film after applying set/reset pulses. The power of primary 785 nm laser is set at low values to prevent crystallization during measurements. As shown in Supplementary Fig.~S7, the normalized Raman spectra for the two randomly chosen conversion cycles exhibit a similar trend; possessing a rather broad peak in the amorphous state and a dual-band peak upon transition to the crystalline state. X-ray photoelectron spectroscopy (XPS) is performed to study the existing elements and determine the binding energies of the core electrons. Core-level spectra of elements are plotted in Supplementary Fig.~S12 obtained through a survey scan within the binding energy range of 0-600 eV. We also leverage X-ray crystallography to determine the atomic structure of GST in its extreme phases. The corresponding X-ray diffraction (XRD) patterns in Supplementary Fig.~S13 show Bragg peaks verifying the face-centered cubic configuration of C-GST. 

\noindent\textbf{In situ electrical characterization.}
In our experiments, full crystallization and amorphization processes are performed by applying a 1.7 V set pulse with 200-$\mu$s-long double exponential waveform and a 3.8 V reset pulse with 200-ns-long rectangular shape, respectively, to the meta-switch. The short pulse used in the latter biasing scheme avoids unwanted material flow during amorphization. Small differences with simulated results are mainly attributed to the discrepancy between the thermal properties of fabricated and simulated materials, the parasitic resistances associated with the probing pads and contacts, random resistance variation of the W patch, and the thermal boundary resistance between contributed materials. Electrical pulses with lower peak voltages than that of the set pulse are also used to transform the state of GST between its extreme phases in multiple states. The voltage pulses features zero width and leading/trailing edge of 100 $\mu$s resolution imposed by the limitations of the source measurement unit (Keithley 2614B). The reset pulse has a leading/trailing edge of $\sim$10 ns that is generated by Tektronix AFG3252C function generator and delivered to ENI 510L RF power amplifier before applying to the device.

\noindent\textbf{In situ optical measurements.}
Experimental optical measurements is performed by directly measuring the intensity of the reflected light from the surface of the fabricated device (see Fig.~2b). A low-power beam (to prevent the conversion of GST during measurements) from a fiber-coupled light source is focused on the device using a 50$\times$ objective lens. To measure the reflected signal a beam splitter is installed right before the surface of the sample to allow separation of the incident and the reflected signals. Normalization is done by dividing the intensity of a reference beam with the same spot size reflected from a smooth surface of an Au patch fabricated near the meta-device under test. To visualize the device under test, a second beam splitter is used in the optical path to direct the reflected visible light to a near-IR charge-coupled device camera. Co-located in situ optical and electrical measurements are carried out while the meta-device under the microscope is connected to the external signal generators with a high frequency Infinity probe.

% Acknowledgements
\medskip
\noindent\textbf{Acknowledgements} \par 
The work was primarily funded by Office of Naval Research (ONR) (N00014-18-1-2055, Dr. B. Bennett) and by Defense Advanced Research
Projects Agency (D19AC00001, Dr. R. Chandrasekar). W.C. acknowledges support from ONR (N00014-17-1-2555) and National Science Foundation (NSF) (DMR-2004749). A. A. (CUNY) acknowledges support from Air Force Office of Scientific Research and the Simons Foundation. M.W. acknowledges support by the Deutsche Forschungsgemeinschaft (SFB 917). M.E.S. acknowledges financial support of NSF-CHE (1608801). This work was performed in part at the Georgia Tech Institute for Electronics and Nanotechnology (IEN), a member of the National Nanotechnology Coordinated Infrastructure (NNCI), which is supported by NSF (ECCS1542174).

% \medskip
% \noindent\textbf{Author contributions} \par 
% S.A. carried out modeling of the devices, conducted device fabrication, and  performed meta-switch characterization. O.H. and H.T. contributed to the device fabrication, material characterization, and multi-state metasurfaces measurements. M.T. and W.C. implemented the optical measurement setup and contributed to the measurements. A.K., A.A.E, and A.A. (CUNY) helped with the optical and electro-thermal analysis. C.T., S.D., E.P., and M.W. optimised and deposited the phase-change material for the metadevices. M.E.S. helped with material characterization. A.A. (GaTech) supervised the whole project. S.A. wrote the manuscript with input from all authors. All authors edited the manuscript.

\medskip
\noindent\textbf{Competing financial interests} \par 
The authors declare no competing interests.

% \medskip
% \noindent\textbf{Supporting Information} \par 
% The online version contains supplementary material available at https://doi.org/

% References
\medskip

% Use the following code if you wish to generate your bibliography with BibTeX;
% replace the string "MSP-template" below with the name(s) of
% the BibTeX data base(s) you want to use.
% The resulting bibliography-output (the content of the .bbl file)
% must be pasted back into this file before submission.
% Please also include your BibTeX data base file(s) in your submission
% so that we can re-run BibTeX if necessary.
%
%\bibliographystyle{MSP}
%\bibliography{MSP-template}
%apsrev4-2.bst 2019-01-14 (MD) hand-edited version of apsrev4-1.bst
%Control: key (0)
%Control: author (8) initials jnrlst
%Control: editor formatted (1) identically to author
%Control: production of article title (0) allowed
%Control: page (0) single
%Control: year (1) truncated
%Control: production of eprint (0) enabled
\providecommand{\noopsort}[1]{}\providecommand{\singleletter}[1]{#1}%
%

% \clearpage
% \newpage

% \end{spacing}

% \end{large}

\end{document}